\documentclass[reprint,aps,prx,superscriptaddress,twocolumn]{revtex4-1}
\usepackage{bm,physics,amsmath,amssymb,
	graphicx,xcolor,embedfile,comment,mathrsfs,mathtools,xfrac}
\usepackage[utf8]{inputenc}
\usepackage[normalem]{ulem}
\usepackage[english]{babel}
\usepackage[T1]{fontenc}
\usepackage{tikz}

\usepackage{titlesec}
\titlespacing*{\section}{0pt}{1\baselineskip}{1\baselineskip}

\graphicspath{{figures/}}

\makeindex

\begin{document}

%-------------------------------------------
% Paper Head
%-------------------------------------------
%\title{Anharmonic Terahertz Polaritonics with Quantum Paraelectric SrTiO$_3$}
\title{Nonlinear Terahertz Polaritonics in a Quantum Paraelectric}
\author{Chao Shen}
\affiliation{ISTA (Institute of Science and Technology Austria),
    Am Campus 1, 3400 Klosterneuburg, Austria}
   
\author{Carla Verdi}
\affiliation{School of Mathematics and Physics, The University of Queensland, Brisbane QLD 4072, Australia}

\author{Serafim Babkin}
\affiliation{ISTA (Institute of Science and Technology Austria),
    Am Campus 1, 3400 Klosterneuburg, Austria}
    
\author{Maksym Serbyn}
\affiliation{ISTA (Institute of Science and Technology Austria),
    Am Campus 1, 3400 Klosterneuburg, Austria}
    
\author{Zhanybek Alpichshev}
\email{alpishev@ist.ac.at}
\affiliation{ISTA (Institute of Science and Technology Austria),
    Am Campus 1, 3400 Klosterneuburg, Austria} 

%\affil[3]{Department, Organization, Street, City, 610101, State, Country}
%\affil[*]{Corresponding author: \texttt{alpishev@ist.ac.at}}

\date{\today}  % Remove date

\begin{abstract}
Terahertz (THz) frequency range holds immense potential for high-speed data processing and signal manipulation. However, a fundamental challenge remains: the efficient and tunable control of THz electromagnetic fields. One promising approach is polaritonic engineering, which leverages hybrid light-matter excitations to manipulate THz fields at sub-wavelength scales. Here, we introduce quantum paraelectric materials as a powerful new platform for THz phonon-polaritonics, leveraging the pronounced nonlinearities of incipient ferroelectrics. These nonlinearities enable strong self- and cross-coupling between polaritons, facilitating all-optical, reconfigurable THz signal control. Using a novel space- and time-resolved imaging technique, we directly observe the ballistic propagation of bulk phonon-polaritons in SrTiO$_3$, and uncover soliton-like, dispersion-free transport in its low-temperature, quantum-fluctuation-dominated phase. Our results establish quantum paraelectric solids as a versatile and highly tunable medium for next-generation THz photonics and ultrafast information processing.
%Terahertz frequency range holds immense potential for high-speed data processing and signal manipulation. However, a fundamental challenge remains: the efficient control of THz-range electromagnetic fields. One promising approach is polaritonic engineering, which leverages hybrid light-matter excitations to manipulate THz fields at sub-wavelength scales. Here, we introduce Quantum Paraelectric (QP) solids as a novel platform for THz phonon-polaritonics, exploiting the extreme nonlinear response inherent to incipient ferroelectrics. This nonlinearity enables strong self- and cross-coupling between polaritons, paving the way for all-optical, field-programmable THz polariton circuits. Using a newly developed space- and time-resolved technique, we directly observe free ballistic propagation of bulk phonon-polaritons in SrTiO$_3$ and reveal dispersion-free, soliton-like polariton transport in the quantum-fluctuation-dominated low-temperature phase. These findings establish QP solids as a robust, highly tunable platform for next-generation THz signal processing and ultrafast optical computing.
\end{abstract}

\maketitle

%\textbf{Keywords}: Quantum paraelectric, Polariton, STO, Soliton, Nonlinear Terahertz spectroscopy.  

%-------------------------------------------
% Paper Body
%-------------------------------------------
%--- Section ---%
 
Fundamental limitations of conventional electronics \cite{Sze2006, Weste2010} imply that the prospects of processing data at the THz rate and beyond hinge on the possibility of controlling electromagnetic signals with light itself. Since electrodynamics is famously linear in vacuum \cite{maxwell1954}, this is only possible inside media where light exists in the hybrid form mixed with matter degrees of freedom and therefore efficiently mediates light-light interactions \cite{Akhieser1936}. According to conventional nonlinear optics, the efficiency of such processes can be improved in two ways: either enhancing the intensities of fields involved or boosting nonlinear susceptibility \cite{Boyd2008}.

The hybrid light-matter modes inside extended matter substance are commonly referred to as polaritons \cite{Pekar1957}. By maximizing the matter fraction one can achieve strong interaction between polaritons \cite{Feurer2007}. A drawback of such straightforward approach is that polaritons with significant matter fraction suffer from substantial chromatic dispersion \cite{Tolpygo1950, Huang1951}, which severely limits their utility for coherent data transmission. An alternative approach to achieve strong coupling between polaritons while keeping dispersion moderate is to enhance the field magnitude in them. To this end, one can confine the field in one spatial dimension - {\it e.g.} through hyperbolic dispersion in the material - while allowing the field to propagate in the orthogonal plane \cite{Basov2016}. This way it is possible to obtain highly nonlinear polaritons at moderate input powers. Unfortunately, the classical wave equation in even-dimensional spaces, generally speaking, does not support transmission of localized pulses \cite{Morley1985}. This observation limits the usefulness of 2D polaritons in the context of information exchange. 

In this work, we demonstrate that the roadblocks to polariton-based optical data processing discussed above can be overcome by accessing the non-perturbative regime that emerges in the low-temperature quantum-critical phase of paraelectric SrTiO$_3$ \cite{Rowley2014}. Specifically, by utilizing phase-sensitive two-dimensional THz Kerr effect spectroscopy to monitor the space- and time-resolved propagation of THz excitations in bulk SrTiO$_3$, we show that the hybridization of a sufficiently intense electromagnetic pulse with the anharmonic soft mode gives rise to a soliton-like phonon-polariton that traverses the crystal with negligible dispersion. These self-confined, long-lived polaritons provide a robust mechanism for information transport and establish a new platform for ultrafast, polariton-based signal processing.

\subsection*{THz Kerr Effect as a probe of bulk-propagating phonon-polaritons}\label{sec1}
\begin{figure*}[htbp!]
    \centering
    \includegraphics[scale = 0.33]{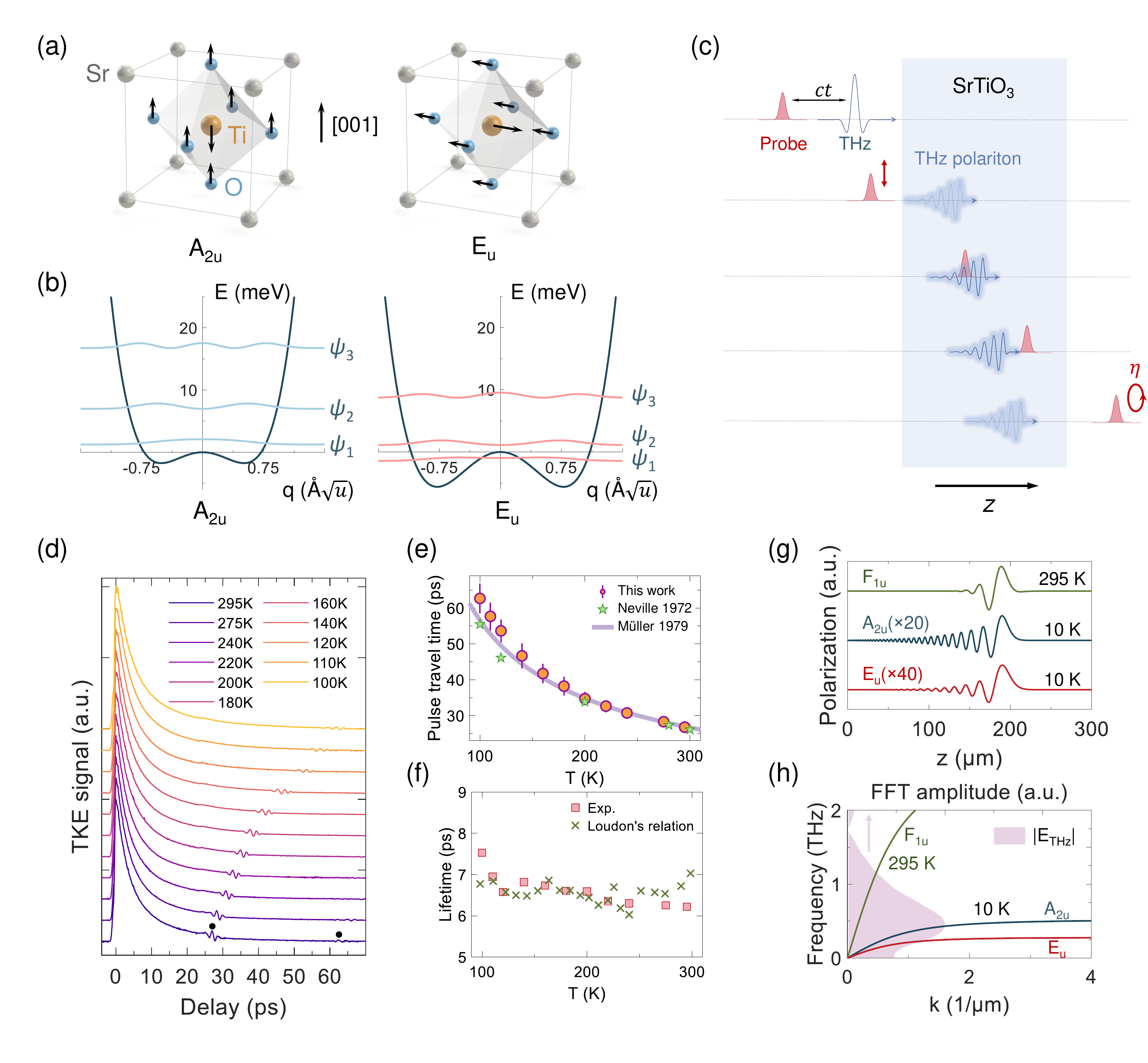}
\caption{\textbf{Soft-mode hybridization with the THz field, experimental schematic, and evidence of propagating polaritons.} (a) Schematic of the $A_{2u}$ and $E_u$ infrared-active phonon modes in SrTiO$_3$. (b) Calculated anharmonic lattice potential of both modes and excited energy levels and wavefunctions solved using a 1D-Schrödinger equation (Methods). The mode amplitude $Q$ is expressed in the unit of $\text{\r{A}}\sqrt{u}$, where $u$ is the atomic mass unit (Dalton). (c) Time-resolved single-pulse TKE measurement illustrated as a series of consecutive time-frames: a linearly polarized optical probe pulse launched with delay $t$, acquires ellipticity $\eta$ due to Kerr interaction with a THz pulse inside the sample. %Owing to the slower group velocity of the THz polariton compared to the probe, the probe samples the full spatiotemporal polarization profile $P(z, t)$}
Note that since the probe is propagating much faster, it catches up with the THz pulse invariably, unless $t$ is very large; the accumulated ellipticity $\eta$ is given by equation (\ref{eq1}). (d) Single-pulse TKE measurements from 295 K down to 100 K, revealing clear echo signals (black dots) at room temperature. At lower temperatures, echoes diminish due to increased THz reflection losses. (e) Comparison between measured (circles) and calculated (stars and solid line) echo arrival times. Markers indicate the average arrival time, and error bars denote the temporal width of the echoes. Calculations use reported dielectric constants of STO \cite{neville1972permittivity, muller1979srti} and account for the finite velocity of the optical probe. (f) Temperature dependence of the polariton lifetime. Experimental values (squares) are extracted via single-exponential fits. Theoretical predictions (crosses) are based on Loudon’s relation. (g) Spatial profiles of the THz polarization field at 295 K and 10 K after 200 $\mu$m of propagation through STO. (h) Calculated dispersion relations of soft-mode polaritons at 295 K and 10 K, overlaid with the measured THz spectrum for comparison.}   
    \label{fig:1}
\end{figure*}

SrTiO$_3$ (STO) is a dielectric perovskite that undergoes a structural phase transition at $T_0 = 105$~K \cite{Vogt1995}, where the triply degenerate $F_{1u}$ phonon splits into an $A_{2u}$ mode (polarized along the $c$-axis) and a doubly degenerate $E_u$ mode (polarized in the $ab$-plane), shown in Fig.~\ref{fig:1}a. Upon further cooling, both modes progressively soften, signaling the approach to ferroelectric instability \cite{Yamada1969, Vogt1995, Yamanaka2000-bg}. However, in contrast to related perovskites like BaTiO$_3$, the growth of the dielectric susceptibility is arrested near $T_c \approx 35$~K by strong quantum fluctuations of lattice ions, which suppress the emergence of long-range ferroelectric order. The resulting ground state is termed Quantum Paraelectric (QP) \cite{muller1979srti}. These soft modes, primarily associated with Ti–O displacements \cite{Cowley1964, Yamada1969}, evolve from being nearly harmonic at high temperature into an intrinsically anharmonic motion in a double-well potential in the QP regime (Fig.~\ref{fig:1}b), reflecting an incipient broken-symmetry state \cite{Zhong1996,carla2023}. The central question we address here is how light interacts with and hybridizes with these anharmonic modes.\\
\indent In the experiment (Fig.~\ref{fig:1}c and see Supplementary Information Section 1), we excite THz phonon-polaritons in STO by irradiating the sample with intense single-cycle THz pulses generated via the tilted-pulse-front technique in a nonlinear crystal. To track polariton dynamics in real time, we use an ultrashort near-infrared probe pulse ($\lambda = 800$ nm) delayed by time $t$, which detects refractive index changes induced by the polariton through the THz Kerr effect (TKE) \cite{hoffmann2009terahertz, finneran2016coherent,maehrlein2021decoding,frenzel2023nonlinear}. Figure \ref{fig:1}d shows the TKE signal versus probe delay at various temperatures above $T_0$. Interestingly, the signal displays relaxor-type dissipative dynamics \cite{li2019terahertz, Cheng2023, li2023terahertz, talbayev2025,blanchard2022two}, with a decay rate (solid marks in Fig.~\ref{fig:1}f) that does not align with known channels like the damping of the \( F_{1u} \) phonon mode at the corresponding temperatures \cite{Vogt1995}. Instead, it originates from the spatial Kerr interaction between the THz polariton and the probe. To see this, we note that while Kerr coupling between propagating non-degenerate pump and probe pulses can get complex in general, it simplifies when the probe’s group velocity greatly exceeds the pump’s. In this limit—valid for THz-pump/near-infrared probe setups in STO—the probe effectively samples a frozen polariton, and the cumulative TKE-induced ellipticity $\eta$ is given by:
\begin{equation}
    \eta(t) \propto \int_0^L \!\!\! dz \,\left( {P}(z,t)\right)^2
\label{eq1}
\end{equation}
\noindent where $L$ is the sample thickness and ${P}(z,t)$ is the local electric polarization of the polariton at spatial coordinate $z$, as seen by the fast probe pulse after delay time $t$. Note that in a medium with large dielectric permittivity $\varepsilon$ the energy of a polariton is stored almost entirely in the mechanical (lattice) degrees of freedom \cite{Brillouin1960, Landau1984}. Since ${P}(z,t)$ is proportional to the displacement of atoms in the lattice, the TKE signal $\eta$ is directly proportional to the potential energy of the polariton stored. Since for a unidirectionally propagating wave the potential energy is always a half of its total energy, the dynamics observed in Fig.~\ref{fig:1}d shows nothing but the real-time decay of a phonon-polariton in STO. Further evidence is provided by the excellent agreement between the decay rates of TKE, and that of the polariton calculated according to Loudon's relation ~\cite{Loudon1970} $\gamma(T) = \gamma_\text{TO}(T) \cdot \left( \,\tilde{\omega}/{\omega_\text{TO}}(T)\right)^2$ (Fig.~\ref{fig:1}e), where $\omega_\text{TO}$ and $\gamma_\text{TO}$ are the resonance frequency and damping rate of \( F_{1u} \) phonon \cite{Vogt1995} and $\tilde{\omega}$ is the characteristic frequency of the polariton pulse.\\
\indent In addition, there is a prominent wave-like feature on the otherwise uniformly decaying TKE signal (see oscillations marked by black dots in Fig.~\ref{fig:1}d). To understand its origin, we note that the interpretation of TKE signal as proportional to the potential and, consequently, total polariton energy is only valid for unidirectional propagation of the latter. This is not always true. In particular, when a wave packet encounters a boundary and interferes with its back-reflection, it cannot be considered unidirectional. In this case, the total energy will alternate periodically between potential and kinetic sectors, manifested as oscillations in the TKE signal (Methods). Note how in Fig.~\ref{fig:1}e these oscillations occur precisely at the moments of time when the polariton is expected to reach the back surface of a 500 $\mu$m-thick STO substrate, providing definitive proof that THz response in STO can be described entirely in terms of propagating polaritons, thus clarifying the qualitative nature of THz response in STO \cite{li2019terahertz, Cheng2023, li2023terahertz, talbayev2025, rubio2024}.\\

\begin{figure*}[htbp!]
    \centering
    \includegraphics[width=0.9\textwidth]{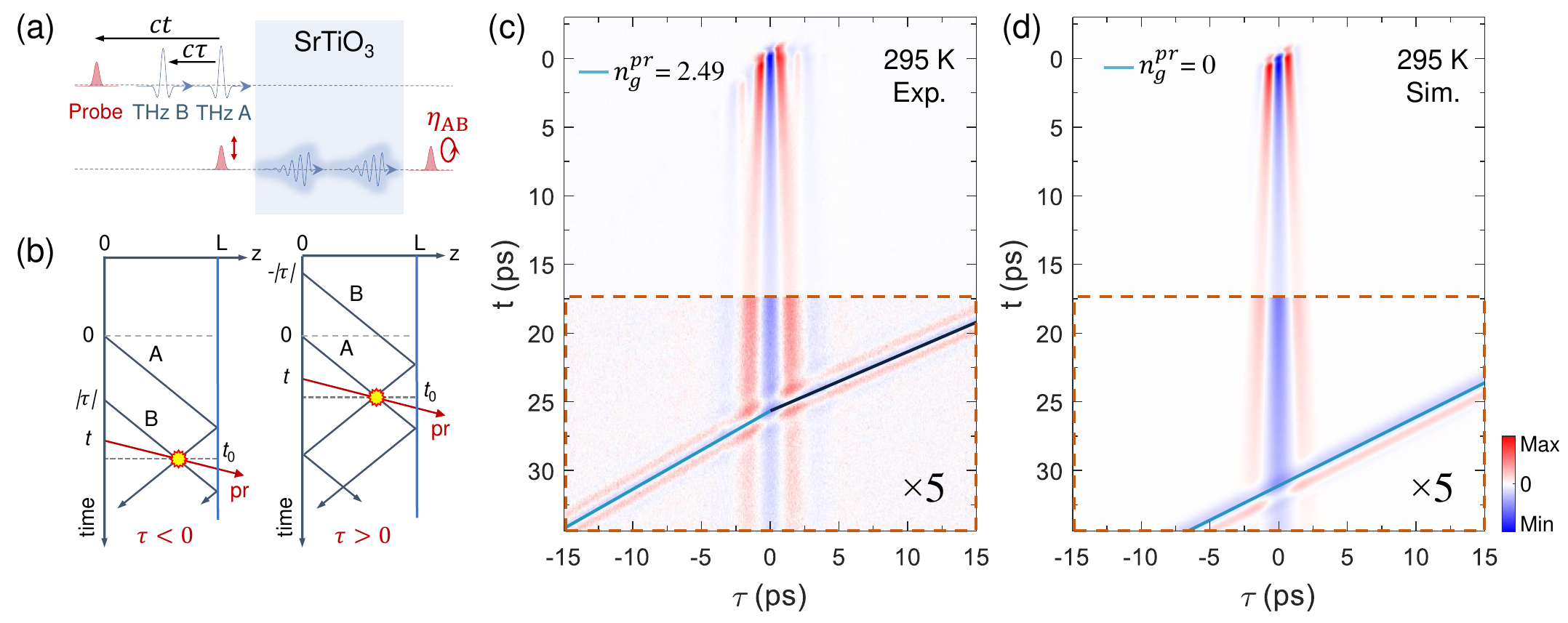}
    \caption{ \textbf{Schematic of 2D-TKE measurement and room temperature 2D-TKE response.} (a) Schematic of the nonlinear 2D-TKE  measurement. Two THz pulses with a controlled delay $\tau$ are incident on bulk STO, and the optical probe samples the birefringence induced by both polaritons. The ellipticity contributions from the individual polaritons are subtracted from the accumulated ellipticity $\eta_\text{AB}$ via a differential chopping scheme to extract the nonlinear signal $\eta_\text{NL}$. (b) World lines illustrating THz and probe propagation in bulk STO. $t$ and $t_0$ denote the $y$-axes of the measured and simulated 2D-TKE maps, respectively. (c) Experimental 2D-TKE response at room temperature. The region within the dashed rectangle is magnified by a factor of 5. Solid lines indicate echo trajectories derived from the world lines, assuming group refractive indices of 2.49 and 17.88 for the probe and THz pulses, respectively. The change in color reflects the slope discontinuity between the $\tau<0$ and $\tau>0$ regimes. (d) Simulated 2D-TKE response using $\eta_\text{NL} \propto \int_0^L  P_\text{A}(z,t)P_\text{B}(z,t+\tau)\,dz$, with phonon frequency $\omega_\text{TO}/2\pi = 2.7$~THz,  damping constant $\gamma_\text{TO}/2\pi = 0.4$~THz and static dielectric constant $\varepsilon_0\approx350$. The region within the dashed rectangle is magnified by a factor of 5. The solid line corresponds to the calculated echo line assuming an infinite probe velocity in the sample.}      
        \label{fig:2}
\end{figure*}
\subsection*{Space-time probing of polaritons via 2D-TKE}\label{sec2}
\indent Apart from confirming the validity of the polariton picture, oscillations in the TKE signal can also be used to reconstruct the polarization field profiles in the vicinity of sample surfaces (Methods). It is possible to generalize this approach and probe the profile of a polariton everywhere, by introducing an auxiliary polariton which can be set to interfere with the original one at an arbitrary location by a suitable choice of timing between them.\\
\indent Experimentally, this can be realized in a 2D-TKE setting whereby two independently generated THz pulses (A and B) launch a pair of polaritons inside the sample at a mutual delay $\tau$ (Fig.~\ref{fig:2}a and see Supplementary Information Section 1). The nonlinear signal $\eta_{\text{NL}}$ corresponding to the interplay between the two pump pulses is defined as the difference between the TKE signal in the presence of both and TKE responses to individual pump pulses:
\begin{equation}
    \begin{split}
        \eta_\text{NL}(t+\tau)\equiv \,\,&\eta\left({P}_\text{A}+{P}_\text{B}\right)-\eta({P}_\text{A})-\eta({P}_\text{B})\\
        \propto &\int_0^L{P}_\text{A}(z,t){P}_\text{B}(z,t+\tau)dz
    \end{split}
    \label{eq:crossed_pol}
\end{equation}
\noindent where $\tau$ is the time delay between pulses A and B \cite{sajadi2015terahertz}. The polarizations of A and B pump pulses are kept aligned with the $c$- and $a$-axes of STO, respectively. This configuration becomes important in the low-temperature tetragonal phase, where A selectively excites the $A_{2u}$ phonon and B excites the $E_{u}$ soft mode.\\
\indent We benchmark this method in the regime where soft mode anharmonicity is weak and Drude-Lorentz model is expected to provide an accurate description of light-matter interaction. In Fig.~\ref{fig:2}c we show the experimental nonlinear 2D-TKE signal $\eta_{\text{NL}}(t,\tau)$ at room temperature and in Fig.~\ref{fig:2}d the same quantity as calculated using equation (\ref{eq:crossed_pol}) with polarizations $P_\text{A}$ and $P_\text{B}$ each obtained by numerically solving the Drude-Lorentz model (Methods). As expected, the agreement between the two is evident, indicating the validity of the harmonic approximation in this regime.\\
\indent Qualitatively, the data in Fig.~\ref{fig:2}c can be understood in terms of equation (\ref{eq:crossed_pol}) with ballistically propagating polaritons. The main feature in Figs.~\ref{fig:2}c and \ref{fig:2}d -- the three-striped structure around $\tau=0$ -- corresponds to the correlation function between $P_\text{A}(z,t)$ and $P_\text{B}(z,t)$ (Methods). The width of the central feature is visibly growing with probe time $t$, which can be understood in terms of chromatic dispersion of the polariton (note the finite curvature of the green curve in Fig.~\ref{fig:1}h). A distinct slanted linear feature appears in the lower half of Fig.~\ref{fig:2}c, marked by a subtle change in slope at $\tau=0$. This corresponds to collision events whereby one polariton meets with another, counter-propagating one reflected from the opposite surface of the sample (see Fig.~\ref{fig:2}b). The simulation in Fig.~\ref{fig:2}d also captures this feature, although with a slightly different location at $\tau=0$ and slope. The origin of the discrepancy lies in the approximate character of equations~(\ref{eq1}) and (\ref{eq:crossed_pol}), which were derived for instantaneous probe ($v_{pr}\rightarrow \infty$).
%(i.e., $n^{pr}_\text{g} \rightarrow 0$)
Substituting instead the experimental values for probe %($n^{\text{pr}}_\text{g}\approx 2.49$)
and THz pump 
%($n^{\text{THz}}_\text{g}\approx17.88$)
pulses, we recover the excellent agreement with data (dark- and light-blue solid lines in Fig.~\ref{fig:2}c; also see Supplementary Information Section 2). 
\begin{figure*}[t!]
    \centering
    \includegraphics[width=0.9\textwidth]{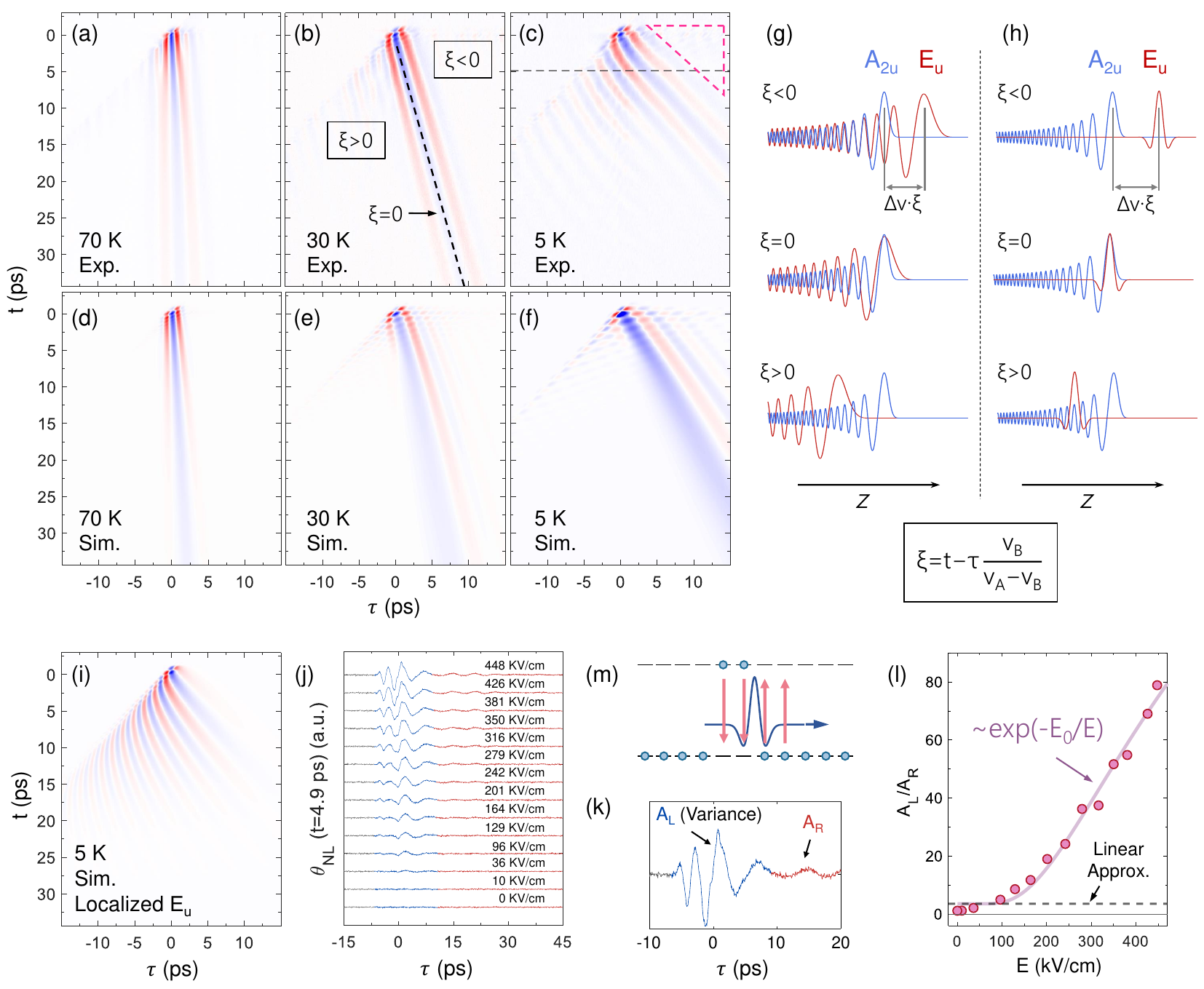}
    \caption{   
    \textbf{Experimental and simulated 2D-TKE responses at low temperatures.} (a–c) Experimental 2D-TKE maps measured at 70, 30, and 5~K. In panel (c), the dashed line marks the $t = 4.9~\text{ps}$ cut, and the magenta triangle indicates the region where nonlinear signal is expected from the Drude–Lorentz model. (d–f) Simulated 2D-TKE maps at 70, 30, and 5~K using the Drude–Lorentz model. The color contrast in panel (f) is adjusted to emphasize the non-empty triangle region corresponding to (c). (g) Schematic illustration of the spatial interaction between dispersed $A_{2u}$ and $E_u$ polaritons, corresponding to different regions in panel (b). Based on which polariton leads in space, the 2D-TKE map is divided into three regimes: $\xi < 0$ ($E_u$ leads), $\xi > 0$ ($A_{2u}$ leads), and $\xi = 0$ (the meeting line), where the $A_{2u}$ polariton overtakes the $E_u$ one after a time delay $t = \tau v_\text{B}/(v_\text{A} - v_\text{B})$. Here, $\xi$ is defined as $\xi = t - \tau v_\text{B}/(v_\text{A} - v_\text{B})$, and the spatial separation between the two polaritons is given by $\Delta v \cdot \xi$, where $\Delta v = |v_\text{B} - v_\text{A}|$. (h) Same as (g), but assuming a localized $E_u$ polariton. (i) Simulation of the 5~K map using a localized $E_u$ polariton, consistent with observations in (c). (j) Nonlinear signal $\eta_\text{NL}$ at $t = 4.9~\text{ps}$ (dashed line in panel (b)) plotted against the electric field strength of pulse B, with pulse A fixed at its maximum. (k) Variance analysis of the $\xi < 0$ region ($A_\text{R}$, red) and the $\xi \ge 0$ region ($A_\text{L}$, blue) as a function of pulse B field strength. As $\eta_\text{NL}$ along the $\tau$ cut is sensitive to the spatial correlation between the two polaritons, $A_\text{L}$ and $A_\text{R}$ track the evolution of the main/leading part and trailing edge of the $E_u$ polariton, respectively. (l) Ratio $A_\text{L}/A_\text{R}$ versus field strength, fitted to $y \sim \exp(-E_0/E)$ with threshold $E_0 \approx 0.6~\text{MV/cm}$. (m) Illustration of self-induced transparency for a single-cycle THz pulse in a two-level system.}
    \label{fig:3}
\end{figure*} 
\subsection*{Anharmonic polariton dynamics in the quantum paraelectric regime}\label{sec3}
As temperature is lowered, the soft mode becomes increasingly anharmonic; so much so that in the QP phase the effective potential profile of the phonon develops multiple minima (\cite{Shin2021, carla2023,Zhong1996}; see Fig.~\ref{fig:1}b). This raises a basic question of how to amend the polariton picture of light-matter interaction to account for it, or whether the picture remains valid at all. In order to reveal the character of THz excitations in STO in the QP phase, we follow the gradual evolution of 2D-TKE as we cool down from room- to cryogenic temperatures.\\
\indent The temperature dependence of the 2D-TKE signal and the Drude-Lorentz simulation are shown in Fig.~\ref{fig:3}a-f. We observe that down to $T\approx70$ K the data can be accurately reproduced within this model, implying that the underlying physics remains qualitatively linear. Unlike high-temperature data in Fig.~\ref{fig:2}c, the central triple-stripe features appear no longer vertical. This is expected, since by the design of the experiment, $P_\text{A}$ and $P_\text{B}$ correspond to different phonon branches ($A_{2u}$ and $E_{u}$ respectively) and therefore propagate with different velocities $v_\text{A} > v_\text{B}$ (Fig.~\ref{fig:1}h). The nonlinear TKE signal $\eta_{\text{NL}}(t, \tau)$ is nonzero only when $P_\text{A}(z,t)$ and $P_\text{B}(z,t)$ overlap (equation (\ref{eq:crossed_pol})). Therefore, if the slower polariton $P_\text{B}$ is launched with a head-start $\tau$, then the faster polariton $P_\text{A}$ will catch up with it only at the ``meeting line'' $t = \tau \,v_B/(v_A-v_B)$ resulting in a tilted central structure in the 2D-TKE signal (Fig.~\ref{fig:3}b and middle panel in~\ref{fig:3}g, also see Supplementary Information Section 2). The agreement between experimental data and simulations indicates the qualitative validity of the harmonic polariton picture in this regime.\\
\indent This, however, is no longer true below $T\lesssim30$ K (see Supplementary Information Section 3), the discrepancy between the predictions of the Drude-Lorentz model and the actual data becoming particularly prominent in the lowest temperature regime $T<10$ K. As Fig.~\ref{fig:3}c shows, the 2D-TKE signal at $T=5$ K comes in the form of broad pattern of stripes with alternating sign, and notably leaves empty the section of the $t-\tau$ plane to the right of the meeting line defined above ($\xi<0$; magenta triangle in Fig.~\ref{fig:3}c). In contrast, the Drude-Lorentz model predicts the opposite behavior (Fig.~\ref{fig:3}f), where non-zero signal is confined to the region right of the meeting line ($\xi>0$), while being negligible to the left of it ($\xi<0$). 
\subsection*{Localization of $E_u$ polariton}
\label{sec5}
\indent In order to understand the behavior of THz excitations in QP STO that gives rise to the nonlinear Kerr signal in Fig.~\ref{fig:3}c, it is instructive to first understand on an intuitive level why harmonic polaritons produce a signal as in Fig.~\ref{fig:3}f. As it turns out, the qualitative structure of the calculated TKE in Fig.~\ref{fig:3}f is not accidental but instead reflects the basic character of harmonic polaritons captured by the Drude-Lorentz analysis. Indeed, according to equation~(\ref{eq:convolution}) the signal $\eta_{\text{NL}}(t, \tau)$ is a correlation of the $E_u$ and $A_{2u}$ polaritons. In the harmonic approximation, the profiles of these polaritons are shown in Fig.~\ref{fig:1}g. The important feature of both pulses is that they are spatially chirped, in agreement with their dispersion relations where the harmonics with the smallest wave vector have the highest velocity (Fig.~\ref{fig:1}h). The dispersion of these polaritons within the harmonic approximation and the spectrum of the exciting THz pulse are shown in Fig.~\ref{fig:1}h and the resulting polariton profiles are illustrated in Fig.~\ref{fig:1}g. Since for any given value $k$ the corresponding spatial harmonic in $A_{2u}$ polariton propagates faster than that in $E_u$, one can only expect appreciable spatial overlap between the two when $A_{2u}$ is lagging behind $E_u$ (Top in Fig.~\ref{fig:3}g). Conversely, when $A_{2u}$ polariton is ahead of $E_u$, the spatial frequencies do not coincide anywhere and the correlation of the two is negligible (Bottom in Fig.~\ref{fig:3}g) naturally accounting for the TKE signal structure in Fig.~\ref{fig:3}f.\\
\indent In this light, the experimental 2D-TKE signal in Fig.~\ref{fig:3}c is very puzzling because its discrepancy with the Drude-Lorentz prediction in Fig.~\ref{fig:3}f cannot be accounted for by simple tweaking of the harmonic model. On the other hand, we note that the 2D-TKE data in Fig.~\ref{fig:3}c can be naturally reproduced if we assume that the polariton associated with the $E_u$ mode propagates as a \textit{localized pulse}, while $A_{2u}$ does so as a usual harmonic dispersive polariton (Fig.~\ref{fig:3}h). Indeed, it is straightforward to see in this case, the TKE signal is automatically zero when the $A_{2u}$ polariton lagging is behind the $E_u$ pulse, and it oscillates as a function of relative lag when the $E_u$ pulse overlaps with the dispersive tail of the $A_{2u}$ polariton. In Fig.~\ref{fig:3}i we show that the qualitative features of the experimental data in Fig.~\ref{fig:3}c can be captured in this scenario. \\
\indent To further investigate this unusual behavior of the $E_u$ polariton, we analyze the dependence of the 2D-TKE signal as a function of field strength of the THz B pulse which pumps the $E_u$ mode, while keeping the intensity of the other pulse constant. In Fig.~\ref{fig:3}j we show field dependence of a representative cut through the 2D-TKE signal at $t=4.9$ ps (dashed horizontal line in Fig.~\ref{fig:3}c). As discussed above, for constant $A_{2u}$ polariton profile the signal magnitude to the right of the meeting line is determined by the amplitude of the dispersive tail of the $E_u$ polariton, while the signal at and to the left of it is largely determined by the main/leading part of the polariton. From this it is evident that in the harmonic regime, the signals in both segments should scale similarly (linearly) with field amplitude - a behavior expected to be broken when nonlinear effects arise. In Fig.~\ref{fig:3}l we plot the ratio of the signal magnitude on the meeting line and to the left of it, $A_\text{L}$, to the one to the right of it, $A_\text{R}$. Here both $A_\text{L}$ and $A_\text{R}$ are calculated as the variances of $\eta_\text{NL}$ in the corresponding ranges of $\tau$ (Fig.~\ref{fig:3}k), the exact limits of which were determined by tracing the evolution of $\eta_\text{NL}$ from room temperature - where the limits of central feature around meeting line are obvious - down to the QP regime (see Supplementary Information Section 4). The behavior shown in Fig.~\ref{fig:3}l clearly exhibits the breakdown of the linear picture, since $A_\text{L}/A_\text{R}$ is evidently not constant as a function of pump field amplitude, but instead exhibits threshold behavior with a characteristic field $E_0 \approx 0.6$ MV/cm.\\
\indent Propagation of $E_u$ polariton in the form of a pulse with minimal dispersion signifies a fundamental failure of the harmonic framework of light-matter interaction in the low-temperature regime of STO and constitutes the central finding of the present work. In order to understand how quantum paraelectricity and associated lattice anharmonicity might lead to solitonic behavior, it is instructive to revisit the excitation spectrum of the $E_u$ phonon mode in Fig.~\ref{fig:1}b (Methods). The two lowest-lying energy levels obtained are detached from the rest of the spectrum. This, in turn, suggests that when considering the coupling of the $E_u$ phonon mode to an electromagnetic wave, it is more appropriate to treat this mode as an ensemble of two-level systems rather than harmonic oscillators. The problem of interaction of electromagnetic field with two-level systems is a well-researched problem allowing for exact solutions \cite{Cohen-Tannoudji}. Specifically, it is known that a pulse propagating through a medium of two-level atoms can form a so-called self-induced transparency (SIT) soliton \cite{McCall1967}, including single-cycle THz pulses \cite{sazonov2020self}, as is the case in the present work (Fig.~\ref{fig:3}m). The interpretation of $E_u$ as an SIT soliton is consistent with the behavior in Fig.~\ref{fig:3}l, including threshold field value ${E}^{th}_0 \sim 1$ MV/cm for STO (Methods).\\
\indent To summarize, we demonstrate that the terahertz response of SrTiO$_3$ can be described in terms of ballistically propagating polaritons whose nature evolves with temperature. While the high-temperature dynamics are well described by harmonic models, the low-temperature regime reveals a distinct, weakly dispersive polaritonic pulse shaped by strong lattice anharmonicity. This behavior, characterized by nonlinear threshold behavior as a function of excitation field magnitude, suggests interpretation in terms of self-induced transparency of a THz pulse propagating through an ensemble of inherently anharmonic phonon modes. {Our results uncover novel regime of coherent signal transport governed by quantum lattice dynamics, opening new avenues for ultrafast THz polaritonics.}
\bibliography{references}
\clearpage
\section*{Methods}
{\bf Experimental details}.
The experiment was conducted using a 70\,fs Ti:Sapphire amplifier (Spectra Physics, Solstice) centered at 800\,nm. 90\% of the output was split equally into two arms and focused into separate LiNbO$_3$ crystals via the tilted-pulse-front technique to generate two intense THz pulses, denoted THz A and THz B. A mechanical delay stage controlled the relative timing between the pulses. The polarizations of the two THz beams were made orthogonal by passing THz A through a 90$^\circ$ periscope. After collimation, both beams were combined using a wire grid polarizer and focused onto a double-side-polished, 500\,$\mu$m-thick [100]-cut SrTiO$_3$ crystal (MSE LLC) with a 4-inch focal length parabolic mirror containing a central through-hole. The THz spot sizes (1/$e^2$) on the sample plane measured with a microbolometer camera (Swiss THz, RIGI) were $740\times630$ $\mu$\text{m}$^2$ and $698\times534$ $\mu$\text{m}$^2$ for THz A and B, respectively (see Supplementary Information Section 1).

The peak electric field was estimated using the relation
\[
E = \sqrt{\frac{2ZW}{\pi r^2 \int g(t)^2 \, dt}}
\]
where $E$ is the electric field, $r$ is the beam radius, $Z$ is the vacuum impedance, $W$ is the pulse energy, and $g(t)$ is the normalized temporal profile. Pulse energies were measured with a pyroelectric detector, and THz waveforms were characterized via electro-optic sampling in a 200\,$\mu$m-thick [110]-cut ZnTe crystal. The peak field strengths of THz A and B at the sample
plane in the air were estimated to be $\sim$310\,kV/cm and $\sim$450\,kV/cm, respectively.

The remaining 10\% of the laser output served as the probe beam for TKE measurements. The probe was focused to a radius of $\sim$200\,$\mu$m (1/$e^2$) and used in a transmission geometry to detect polarization ellipticity $\eta$. A differential chopping scheme (see Supplementary Information Section 1) was employed with a lock-in amplifier (SR810) to isolate the nonlinear response. By setting a $\pi/2$ phase shift between the 250\,Hz THz chopping frequency and the reference, the extracted nonlinear Kerr signal is given by
\begin{equation}
\begin{split}
    \eta_{\mathrm{NL}} &= -(\eta_\text{AB}+\eta_\text{G}) + (\eta_\text{A}+\eta_\text{G}) + (\eta_\text{B}+\eta_\text{G}) - \eta_\text{G} \\
    &\propto \eta_\text{AB} - \eta_\text{A} - \eta_\text{B}
\end{split}
\end{equation}
where $\eta_\text{AB}$ is the signal with both THz pulses present, $\eta_\text{A}$ and $\eta_\text{B}$ correspond to single-pulse excitations, and $\eta_\text{G}$ is the static background.\\
{\bf Numerical simulations.} The spatio-temporal polarization field profiles are numerically calculated by solving Maxwell's equations using the Drude-Lorentz dispersion relation in COMSOL. In the Lorentz model, the dielectric function is 
\begin{equation}
    \epsilon(\omega) = \epsilon_{\infty}-\frac{\omega_0^2(\epsilon_0-\epsilon_{\infty})}{\omega^2-\omega_{0}^2+i\Gamma_{0}\omega}
\end{equation}
where $\epsilon_0$ and $\epsilon_{\infty}$ are the dielectric constants in the low- and high- frequency limit, $\omega_0$ and $\Gamma_0$ are the frequency and damping constant of the soft-mode phonon.  Electro-optical sampling waveforms were used as the input pulse and the sample geometry was set to be the same as the experiment. The temperature-dependent phonon frequencies and damping parameters were adapted from Ref. \cite{Vogt1995} and low frequency dielectric constants $\varepsilon_0$ were adapted from Ref.~\cite{Sakudo1971-wq}. The dielectric constant in the optical frequency range was chosen for $\varepsilon_{\infty}$. \\
{\bf Individual conservation of kinetic and potential energies in unidirectionally propagating wave and its breakdown for a general solution.} A general linear medium is characterized by a field of generalized coordinates $u({\bf r},t)$ with conjugate momenta $\pi({\bf r},t)$. The Hamiltonian (after suitable normalization) can be written as 

\begin{equation}
H = \frac{1}{2}\int_V d^3{\bf r} \left\{ u({\bf r},t)^2 +\pi({\bf r},t)^2 \right\} 
\end{equation}

\noindent The total energy can be separated into potential and kinetic energies as 

$$U = \frac{1}{2}\int_V d^3{\bf r} \,u^2 \,\,\,\,\mathrm{and} \,\,\,\,   T = \frac{1}{2}\int_V d^3{\bf r} \,\pi^2$$ 
%Recalling Hamilton's equations
%$$\dot{\pi}({\bf r},t) = -\frac{\delta H}{\delta u} = -u({\bf r},t) \,\,\,\,\mathrm{and} \,\,\,\,\dot{u}({\bf r},t) = \frac{\delta H}{\delta \pi} = \pi({\bf r},t)$$ 
\noindent Now, for solutions propagating in the direction set by a unit vector ${\bf n}$, $u({\bf r},t) = f({\bf r}\cdot {\bf n}-ct)$ and $\pi({\bf r},t)=g({\bf r}\cdot {\bf n}-ct)$, where $c$ is phase velocity and $f(x)$ and $g(x)$ are some arbitrary functions (we chose right-going waves for concreteness), therefore $\dot{u}=-{{\bf \nabla} u}\cdot {\bf n}$ (and same for $\pi$). With this, we arrive at:

\begin{equation}
\begin{split}
 \frac{dU}{dt} = &\int_V d^3{\bf r} \,u \dot{u} = -\int_V d^3{\bf r} \,u \, \left( {\mathbf{\nabla} } u\cdot {\bf n} \right) = \\
= -&\frac{1}{2}\int_V d^3{\bf r} \,\,{\bf n} \cdot \mathbf{\nabla} u^2 = -\frac{1}{2}\int_S dS \, u^2 (\mathbf{n} \cdot \hat{\mathbf{n}}_S)   
\end{split}
\label{eq:en_flux}
\end{equation}
\noindent where $S$ is the boundary of the integration region $V$ and $\hat{\mathbf{n}}_S$ is the vector normal to the unit area $dS$. A similar expression can be produced for $\dot{T}$ in terms of $\pi({\bf r},t)$. \\
%However, for the important case of monochromatic wave of frequency $\nu$ 
%$$\int dt \,\dot{U} \leq \frac{\mathrm{max}(\dot{U})}{2 \nu} \equiv \Delta U$$ 
%\noindent If the length of the propagation region $L \gg c\nu$, then $U \gg \Delta U$ and the surface flux term in eq.\ref{eq:en_flux} can be neglected. Similar considerations yield $\dot{T}=0$. 
\indent The last term in equation (\ref{eq:en_flux}) describes the energy flux in- and out of the integration region and is not equal to zero in general, even for a plane wave in an open region. However, in practice, when it comes to wave propagation in extended media, this is not important, as one typically deals with solutions in the form of pulses. In this case, the surface term in equation (\ref{eq:en_flux}) is zero or negligible for all times except when the pulse is entering or leaving the medium. \\
%formal way to get rid Formally, the surface term can be made identically zero for an arbitrary (unidirectional) solution by imposing periodic boundary conditions on the problem.
\indent We conclude that under the said circumstances, kinetic energy \( T \) and potential energy \( U \) are conserved individually ($\dot{U}=\dot{T}=0$) and can change only due to energy flux through the boundaries. This holds for any solution of the wave equation, provided it is propagating in a single direction, i.e., it can be expressed in the form \( u({\bf r},t) = u({\bf r} \cdot {\bf n} - ct) \). In particular, this is always true for a pulse away from sample edges or an arbitrary wave under periodic boundary conditions. However, in general, this picture breaks down when \( u({\bf r},t) \) is a superposition of more than one non-co-propagating wave solution. In this case only the total energy is conserved, ``sloshing'' back and forth between kinetic and potential sectors.  \\
\indent Note that in the specific case discussed in the main text, one might naively assume that since the enormous refractive index in SrTiO$_3$ ensures that the field on the STO-air interface is virtually zero, the surface term in equation (\ref{eq:en_flux}) vanishes and there is no exchange between kinetic and potential energies near the back surface. This, however, is not true as most of the wave is reflected backwards here due to strong Fresnel reflection, the total field is a superposition of the incoming and outgoing waves, {\it i.e.}, $P\left({\bf r},t\right) \neq P({\bf r} \cdot {\bf n} - ct)$.\\
{\bf Reconstruction of polariton field profile from undulatory feature in TKE response near sample surface.} When a polariton $P(z,t)$ reaches the inner surface of the sample, an  interference between the incident and reflected parts gives rise to an oscillating (undulatory) feature in the TKE signal $\eta(t)$. Here we demonstrate that these oscillations can be used to reconstruct $P(z,t)$ near sample edges. \\
\indent First, we note that according to Fresnel equations, in a material like STO with large refractive index $n$, the total field near the interface can be written as:
\begin{equation}
    P(z,t) \approx f(z,t)+f(-z,t)
\end{equation}
\noindent where $f(z,t)$ is the incident field. Next, we note that if dispersion can be neglected at least for the duration of time that the pulse $f(z,t)$ spends near the interface, then $f(z,t) \approx f(z-c_p t)$ and:
\begin{equation}
\begin{split}
    P(z,t) \approx &f(z-c_p t)+f(-z-c_pt)=\\
    =&\int \frac{d\omega}{2\pi} f_\omega \left[  e^{i \frac{\omega}{c_p}(z-c_p t)}  + e^{i \frac{\omega}{c_p}(-z-c_p t)}  \right] \\
    = &2\int \frac{d\omega}{2\pi} f_\omega e^{-i\omega t} \cos\left( \frac{\omega z}{c_p}\right)
\end{split}
\end{equation}
\noindent from which we obtain for the TKE signal:

\begin{equation}
    \begin{split}
        &\eta(t) \propto \int\limits_{-\infty}^{\infty}dz \,\left( P(z,t) \right)^2=\\
        &=4 \int \frac{d\omega_1}{2\pi} \frac{d\omega_2}{2\pi} f_{\omega_1} f_{\omega_2} e^{-i(\omega_1+\omega_2)t}\times\\
        &\qquad \times \int dz \cos\left( \frac{\omega_1 z}{c_p}\right)\cos\left( \frac{\omega_2 z}{c_p}\right)\\
        &=2\int \frac{d\omega_1}{2\pi} \frac{d\omega_2}{2\pi} f_{\omega_1} f_{\omega_2} e^{-i(\omega_1+\omega_2)t} \times \\
        &\qquad \times \int dz  \left\{ \cos\left[ (\omega_1-\omega_2)\frac{z}{c_p}\right] +\cos\left[ (\omega_1+\omega_2)\frac{z}{c_p}\right] \right\}\\
        &=4\pi c_p\int \frac{d\omega_1}{2\pi} \frac{d\omega_2}{2\pi} f_{\omega_1} f_{\omega_2} e^{-i(\omega_1+\omega_2)t} \times\\
        &\qquad \qquad \times\left[ \delta(\omega_1+\omega_2) +\delta(\omega_1-\omega_2) \right]=\\
        &=2c_p \int \frac{d \omega}{2 \pi} \left[  f_\omega f_\omega e^{-i2\omega t} + f_\omega f_{-\omega}\right]
    \end{split}
    \label{eq:tke_to_f}
\end{equation}

\noindent Here in the second term in square brackets we recognize the non-oscillatory component of the TKE signal, while the first one corresponds to the undulatory feature $\eta_\text{osc}(t)$. Knowing $\eta_\text{osc}(t)$ from experiment, one can obtain $f_\omega$ from equation (\ref{eq:tke_to_f}) as:

\begin{equation}
    (f_\omega)^2 =\frac{1}{c_p} \eta_\text{osc}(2\omega) 
\end{equation}

\noindent where $\eta_\text{osc}(\omega)=\int dt \, e^{i\omega t}\, \eta_\text{osc}(t)$. Note that here we have a square of $f_\omega$ itself, not $|f_\omega|^2$. This means that apart from the overall sign, the phase information on $f_\omega$ is not lost and $f(z,t) \approx f(z-c_p t) =\int \frac{d \omega}{2\pi} f_\omega \exp\left( i \frac{\omega}{c_p} (z-c_p t) \right)$ can be directly reconstructed from the experimentally measured quantity $\eta_\text{osc}(t)$ alone (up to sign).\\  
\noindent {\bf Simplification of $\eta_\text{NL}$ under linear dispersion.} At elevated temperatures the bandwidth of the incident THz pulses lies entirely within the linear part of polariton dispersion (the phonon frequency is $\omega_{\text{TO}}/2\pi\sim2.7$ THz; see Fig.~\ref{fig:1}h), therefore $P(z,t+\tau) \approx P(z-c_{p}\tau, t)$, $c_p$ standing for phase velocity of polariton, and $\eta_\text{NL}(t+\tau)$ can be reduced to spatial correlation between the $P_\text{A}(z,t)$ and $P_\text{B}(z,t)$ offset by $\delta z_\tau =c_p \tau$: 

\begin{equation}
\eta_{\text{NL}}(t, \tau) \propto \int dz P_\text{A}(z,t)P_\text{B}(z-\delta z_\tau,t)
\label{eq:convolution}
\end{equation}

\noindent {\bf Estimating threshold field for SIT.} According to Ref.~\cite{sazonov2020self}, the threshold field is given by $d_{01} E_0 \sim h/\tau_p$, where $h$ is the Planck's constant, $d_{01}$ is the dipole moment of the transition between the ground- and first excited states of $E_u$ phonon, and $\tau_p$ is the duration of the THz pulse. The dipole moment can be estimated from the known material properties as:

$$d_{01} \approx \sqrt{ \frac{\hbar \omega_0 \varepsilon a^3}{4 \pi} }$$

\noindent where $\hbar \omega_0$ is the energy spacing between the two lowest energy levels of the phonon mode; $\varepsilon$ is the dielectric permittivity (at $\omega \sim \omega_0$); and $a$ is the lattice constant of STO. Taking $\tau_p \approx 2$ ps ($\hbar \omega_0 \sim h/\tau_p$) we arrive at:

$${E_0} \sim 10~\mathrm{kV/cm}$$

This should be understood as the threshold value for the field {\it inside} the medium. Accounting for Fresnel losses at the interface of STO ($n\approx100$ in the QP regime), we finally obtain the threshold {\it incident} field:

$$\mathscr{E}_0 \sim 1~\mathrm{MV/cm}$$

\noindent in good agreement with observations.\\
\noindent {\bf Anharmonic lattice potential calculations.} We used the machine-learned force field (MLFF) trained on random-phase approximation (RPA) reference data from Ref.~\cite{carla2023} to compute the anharmonic potential energy surface of the $A_{2u}$ and $E_u$ soft modes (Fig.~\ref{fig:1}b). The calculations were performed using VASP~\cite{Kresse1996,Kresse1999} for tetragonal STO at the experimental low-temperature lattice parameters~\cite{Okazaki1973}, allowing the internal atomic positions to relax in the unit-cell tetragonal geometry until forces were lower than 0.5~meV/\AA. We then calculated the soft-mode phonon frequencies using finite differences. The anharmonic potential energy surfaces were obtained by displacing each atom along the eigenvectors of the imaginary phonon modes. As shown in Ref.~\cite{carla2023}, this approach provides an accurate description of the anharmonic energy landscape in STO, leveraging the efficiency of the MLFF to accelerate RPA calculations~\cite{Liu2022}.\\
\noindent {\bf Energy levels in anharmonic lattice potential.}
To calculate the excited energy levels corresponding to the $A_{2u}$ and $E_u$ modes, we begin by considering the normal modes of the lattice. Each mode is characterized by an effective mass $m$, a displacement $\xi$, and an effective potential $V(\xi)$. In a conventional case, the harmonic approximation can be used for the potential, however, in the case of $A_{2u}$ and $E_u$ modes, it becomes unstable because $\partial^2_\xi V(\xi)\Big|_{\xi=0}<0$, indicating a double-well structure. This instability was revealed by displacing atoms along the soft-mode eigenvectors and evaluating the potential energy surface using first-principles calculations for the values of $c/a= 1.001$, which $c/a$ indicates the tetragonal distortion. The calculated energy profiles (see Fig.~\ref{fig:1}b) show two shallow wells that reflect the anharmonic character of the ferroelectric instabilities in STO.
We obtain the excited energy levels of $A_{2u}$ and $E_u$ modes from the potentials in Fig.~\ref{fig:1}b, by solving the effective one-dimensional Schrödinger equation for each mode. To this end, we move to mass-weighted coordinates, $q=\sqrt m \xi$, which removes the ambiguity in the separate definition of $\xi$ and $m$ \cite{Esswein2022}. Then, the Schrodinger equation takes the form,
\begin{equation}
\left[-\frac{\hbar^{2}}{2}\frac{d^{2}}{dq^{2}}+V\left(q\right)\right]\psi(q)=E\psi(q)
\end{equation}
with the potential given by $V=V_0\left(q^4/\sigma^4-2q^2/\sigma^2+1\right)$, where values of $V_0$ and $\sigma$ are determined from fitting the ab-initio potential with a quartic polynomial. We solve this equation numerically using the effective potentials for $A_{2u}$ and $E_u$, and obtain the excited energy levels (see Fig.~\ref{fig:1}b). From this solution, we observe that the low-lying levels of the $A_{2u}$ mode are close to being equidistant, with splittings $E^{A_{2u}}_2-E^{A_{2u}}_1 = 5.73\, {\rm meV}$ and $E^{A_{2u}}_{3}-E^{A_{2u}}_2 = 9.76\, {\rm meV}$. In contrast, the $E_u$ mode, as expected from a more anharmonic potential, exhibits greater variation between splittings, with $E^{E_{u}}_{2}-E^{E_{u}}_1 = 2.74\, {\rm meV}$ and $E^{E_{u}}_{3}-E^{E_{u}}_2 = 7.62\, {\rm meV} $, which is consistent with the nonlinear behavior observed in the $E_u$ polaritons.\\

\section*{Acknowledgements}\label{sec7}
Z.A. and M.S. acknowledge support from the collaborative research project SFB Q-M\&S funded by the Austrian Science Fund (FWF, grant No.PR1050F86). C.V. acknowledges financial support from the Australian Research Council (DE220101147) and computational resources 
provided by the Australian National Computational Infrastructure and Pawsey Supercomputing Research Centre through the National Computational Merit Allocation Scheme.

\end{document}